\documentclass[onecollarge,natbib]{svjour2}
\bibpunct{[}{]}{;}{n}{}{,} 
\smartqed  
\usepackage{graphicx}
\DeclareMathAlphabet{\mathpzc}{OT1}{pzc}{m}{it} 
\journalname{Few-Body Systems}
\begin{document}

\title{Flavor analysis of nucleon, $\Delta$, and hyperon\\
electromagnetic form factors
}

\author{Martin Rohrmoser \and Ki-Seok Choi \and Willibald Plessas}

\institute{Martin Rohrmoser \at
           SUBATECH, Ecole des Mines de Nantes, F-444307 Nantes, Cedex 3, France
           \and Ki-Seok Choi \at
           Department of Physics, Soongsil University, Seoul 156-743, Republic of Korea
           \and Willibald Plessas \at
              Institute of Physics, University of Graz, A-8010 Graz, Austria \\
              \email{plessas@uni-graz.at}          
           }

\date{Received: date / Accepted: date}

\maketitle

\begin{abstract}
By the analysis of the world data base of elastic electron scattering on the proton and
the neutron (for the latter, in fact, on $^2H$ and $^3He$) important experimental insights
have recently been gained into the flavor compositions of nucleon electromagnetic form
factors. We report on testing the Graz Goldstone-boson-exchange relativistic
constituent-quark model in comparison to the flavor contents in low-energy nucleons, as
revealed from electron-scattering phenomenology. It is found that a satisfactory agreement
is achieved between theory and experiment for momentum transfers up to $Q^2\sim$ 4 GeV$^2$, relying on three-quark configurations only. Analogous studies have been extended to the
$\Delta$ and the hyperon electromagnetic form factors. For them we here show only some
sample results in comparison to data from lattice quantum chromodynamics.

\end{abstract}
\vspace{4mm}

Evidently, electromagnetic (e.m.) form factors provide stringent tests on any model for hadrons.
The Goldstone-boson-exchange (GBE) relativistic constituent-quark model (RCQM) for baryons constructed by the Graz group~\cite{Glozman:1997ag} had been tested with respect to covariant predictions for the elastic e.m. $N$ form factors long ago~\cite{Wagenbrunn:2000es,Boffi:2001zb}. An unprecedented overall agreement with experimental data up to momentum transfers of
$Q^2\sim$ 4 GeV$^2$ had then been achieved in a calculation along point-form relativistic
quantum mechanics. After the appearance of phenomenological flavor analyses of elastic e.m. $N$
form factors~\cite{Cates:2011pz,Qattan:2012zf,Diehl:2013xca} it appeared more than interesting
to check the performance of the GBE RCQM also in these respects. Recently we have performed
such studies. Below we show pertinent results of selective quantities for the $N$ and from
extensions of this kind of investigations to the $\Delta$ and to the hyperons with various
$u$, $d$, and $s$ quark contents.
 
The theory and the calculations are exactly the same as explained for the point-form approach
in our previous papers~\cite{Wagenbrunn:2000es,Boffi:2001zb,Melde:2004qu,Melde:2007zz}. The
predictions for the elastic e.m. form factors fulfill Poincar\'e invariance as well as
time-reversal invariance and current conservation~\cite{Melde:2004qu,Melde:2007zz}.
Accurate three-quark baryon wave functions were obtained
solving a relativistically invariant mass operator along the stochastic variational method
exploiting all possible symmetries in configuration, spin, and flavor spaces. For the rest
frames they are depicted in Ref.~\cite{Melde:2008yr} for singlet, octet and decuplet baryon
states. In the calculation of the e.m. form factors the necessary Lorentz boosts can be
executed rigorously when evaluating the matrix elements of the e.m. current operator in the framework of the point form.

\begin{figure}[t]
\begin{center}
\includegraphics[width=0.49\textwidth]{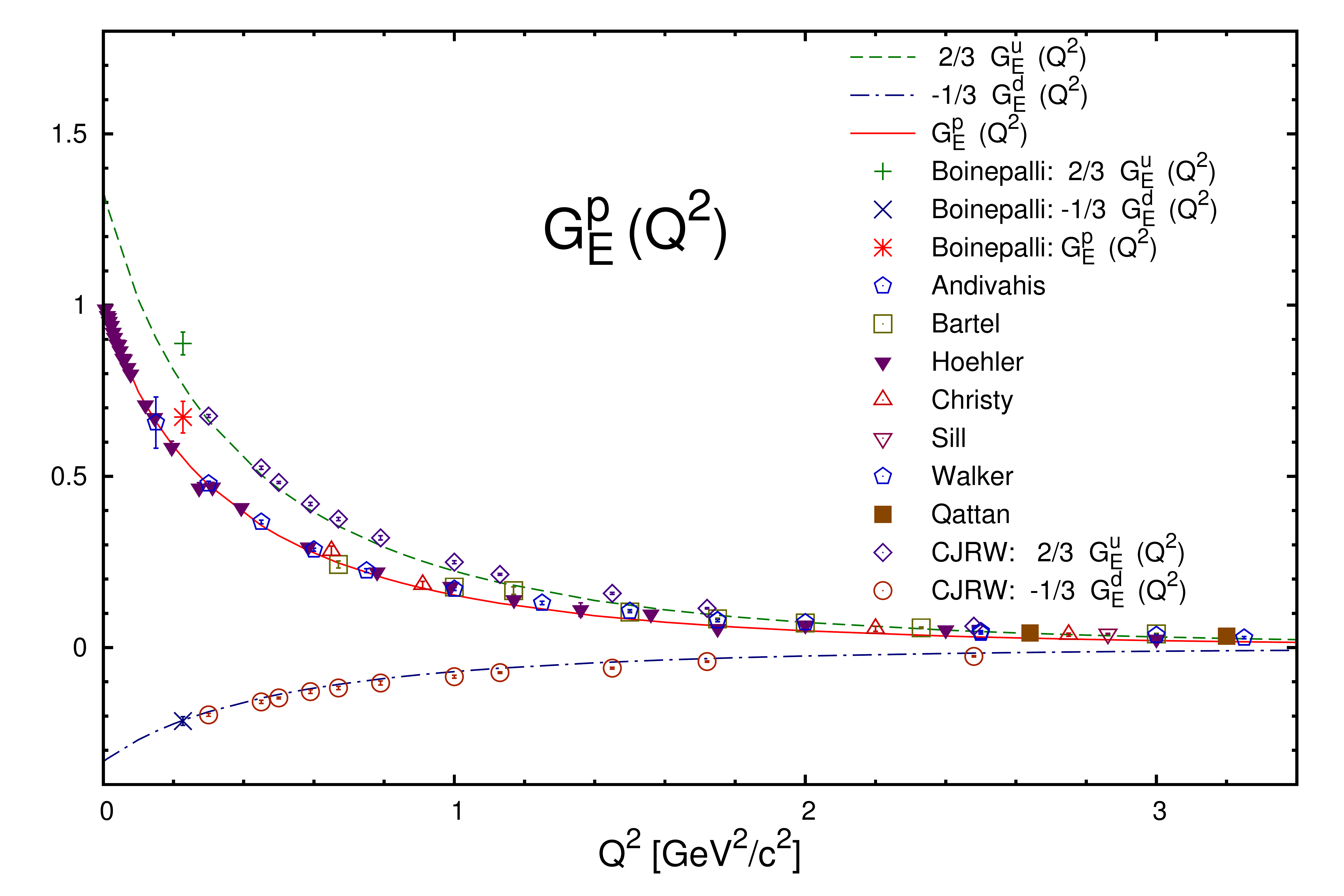}
\includegraphics[width=0.49\textwidth]{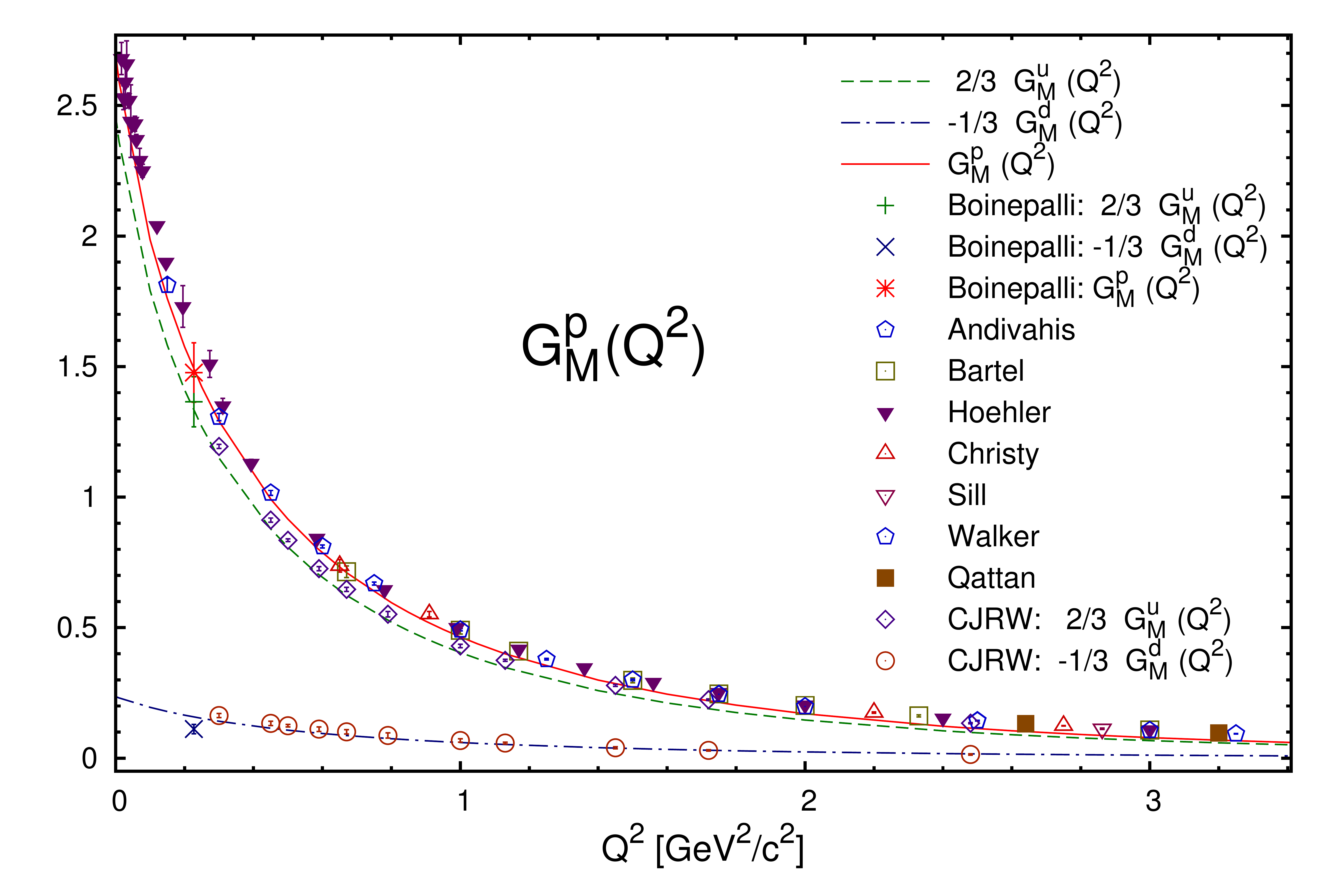}\\[3mm]
\includegraphics[width=0.49\textwidth]{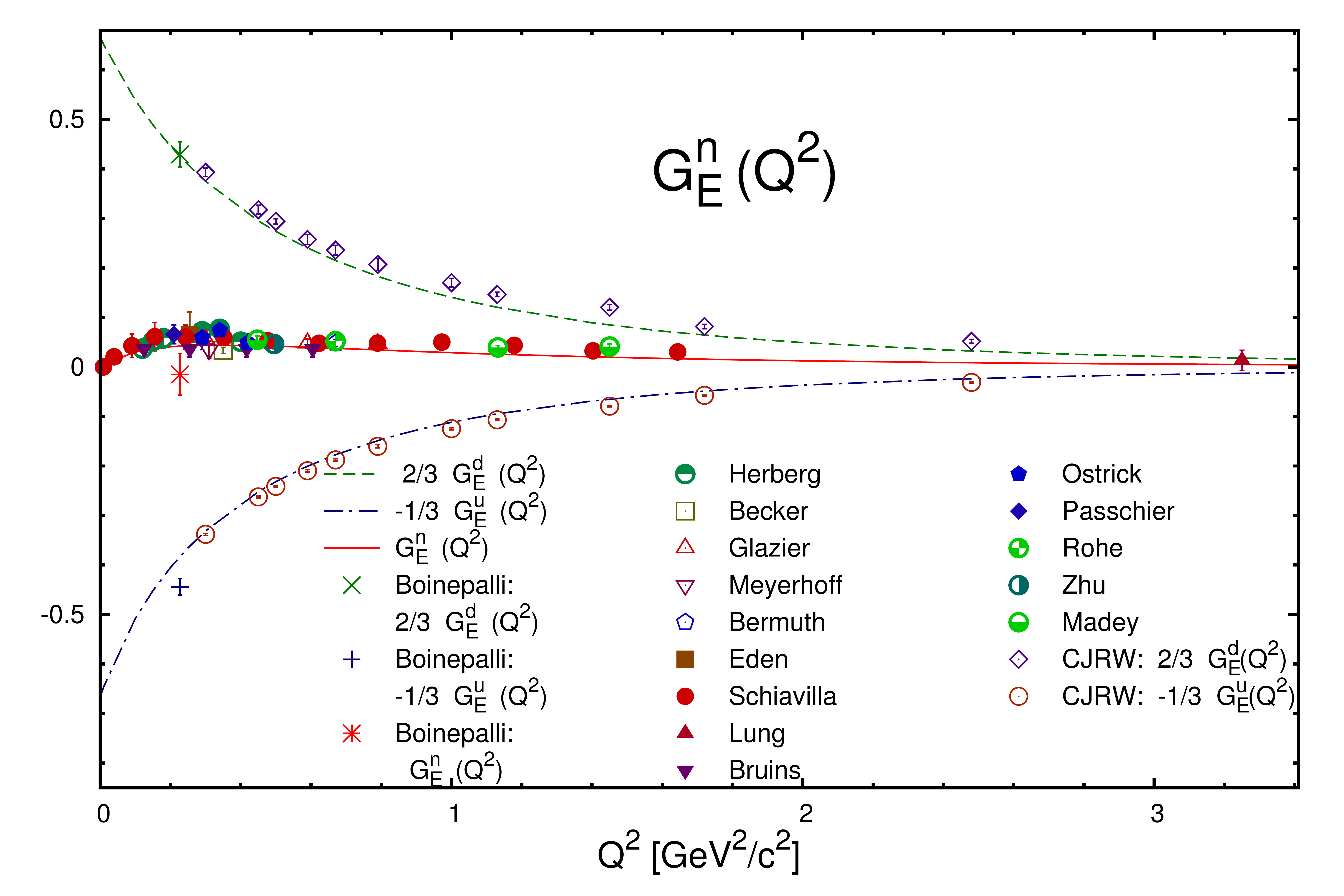}
\includegraphics[width=0.49\textwidth]{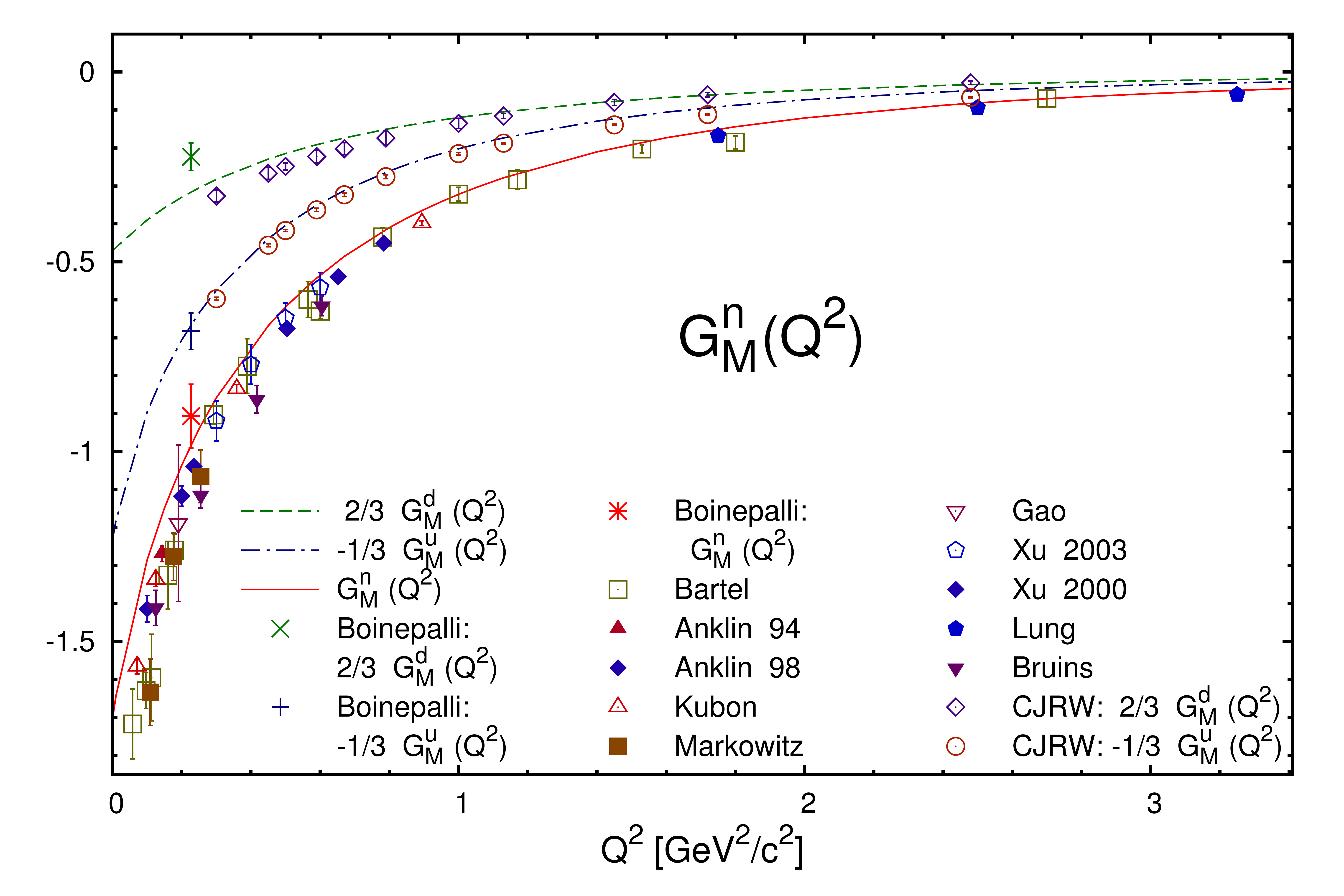}
\end{center}
\caption{Covariant predictions of the GBE RCQM for electric and magnetic Sachs form factors
of $p$ and $n$ (solid/red lines) in comparison to available data from $e^--$scattering.
The various flavor components (broken lines as specified in the inserts) are compared to
the phenomenological data by Cates et al.~\cite{Cates:2011pz} (CJRW) and to a lattice QCD calculation by Boinepalli et al.~\cite{Boinepalli:2006xd}.} 
\label{NFFs}
\end{figure}

In Fig.~\ref{NFFs} we first show the e.m. form factors of both the proton ($p$) and
neutron ($n$) together with their $u$- and $d$-flavor
components $G^u_E$, $G^d_E$, $G^u_M$, and $G^d_M$. It is seen that not only the global predictions by the GBE RCQM agree remarkably well with experimental data but also the
individual flavor contributions are quite congruent with the phenomenological analysis by
Cates et al.~\cite{Cates:2011pz}. For the particular value of $Q^2$=0.227$\pm$0.002 GeV$^2$
there is also a result available from lattice quantum chromodynamics
(QCD)~\cite{Boinepalli:2006xd}, which we quote too in Fig.~\ref{NFFs}.

\begin{figure}[h]
\begin{center}
\includegraphics[width=0.49\textwidth]{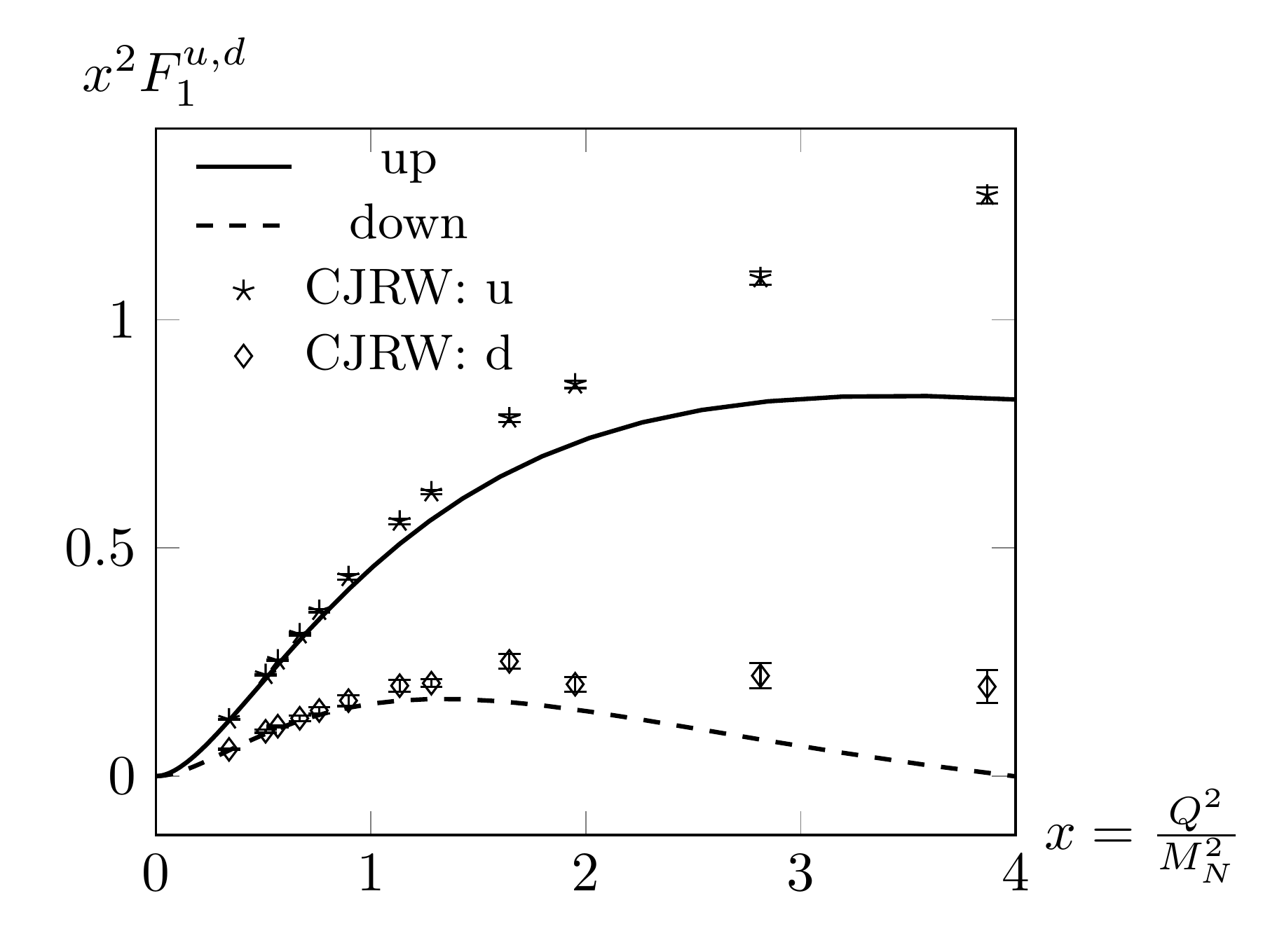}
\includegraphics[width=0.49\textwidth]{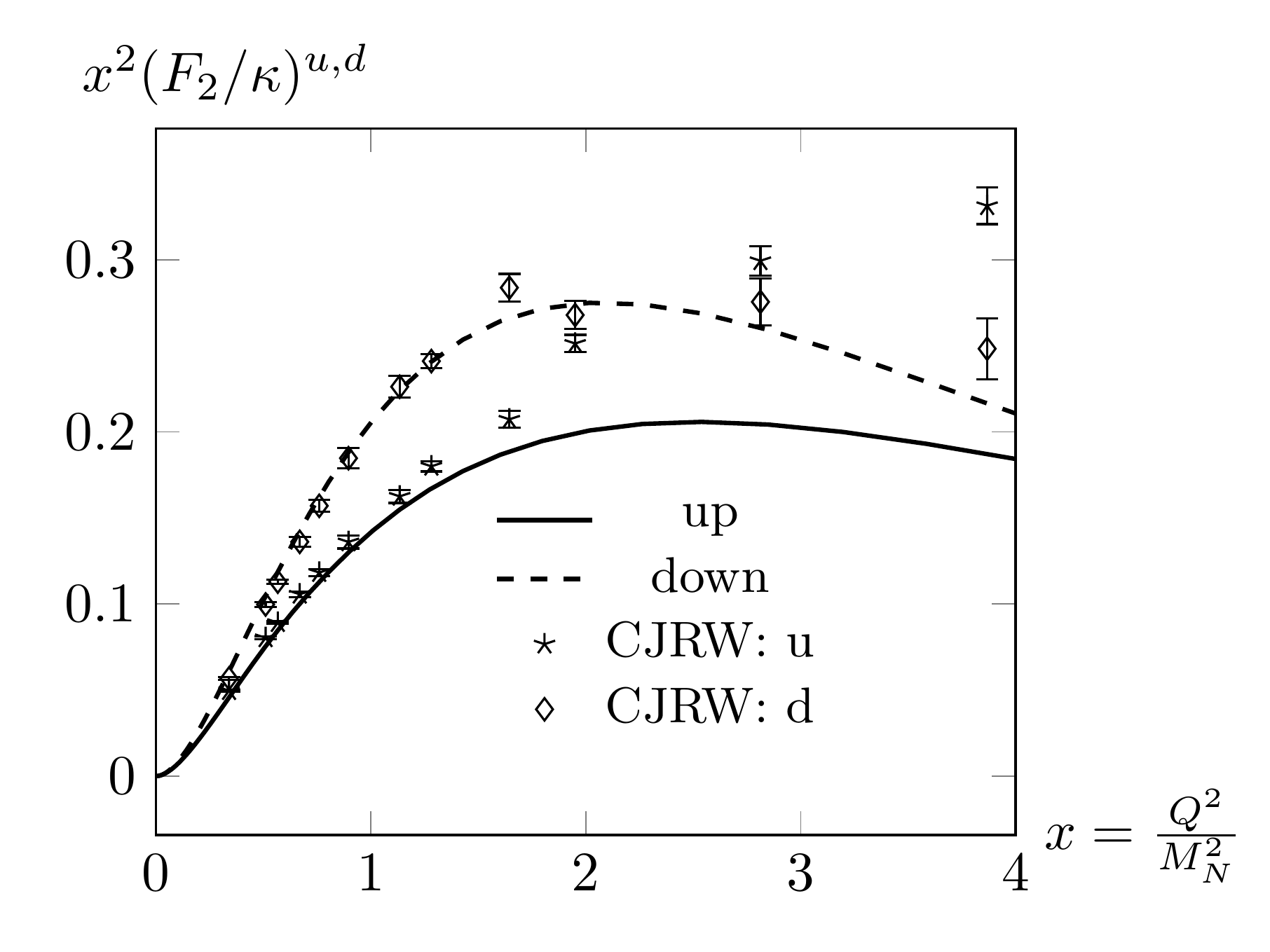}
\end{center}
\caption{$u$- and $d$-flavor contributions to the Dirac and Pauli $N$ form factors
$F_1(Q^2)$ and $F_2(Q^2)$, respectively,
as produced by the GBE RCQM (solid and dashed lines as specified in the inserts). In case of $F_2^{u,d}$ we have plotted the ratio by the corresponding contributions
$\kappa^{u,d}$ to the $p$ and $n$ anomalous magnetic moments. The comparison is made to the phenomenological data by Cates et al.~\cite{Cates:2011pz} (stars and diamonds).} 
\label{NFFratios}
\end{figure}

Sometimes the $N$ flavor components are represented also by flavor contributions to
the Dirac and Pauli form factors $F_1(Q^2)$ and $F_2(Q^2)$, respectively.
For the sake of comparison with
other studies in the literature we add in Fig.~\ref{NFFratios} the $u$- and $d$-flavor components
$F_1^{u}$, $F_1^{d}$, $F_2^{u}$, and $F_2^{d}$ again in comparison to the phenomenological data
by Cates et al. It is seen that the GBE RCQM relying on $\{QQQ\}$ configurations only can well
produce the magnitudes and shapes of all of these form-factor components. The slight deviations
from the phenomenological data at $Q^2\ge$ 3 GeV$^2$ in our opinion do not allow to draw
conclusions of diquark clustering or higher quark Fock components in the low-energy $N$,
as is sometimes advocated in the literature.

Of course, the GBE RCQM could be fine-tuned to produce an even better description of the $N$
e.m. form factors as well as electric radii and magnetic moments
(cf. Refs.~\cite{Wagenbrunn:2000es,Boffi:2001zb,Berger:2004yi}). We emphasize here
again that beyond the definition of the GBE RCQM no further parameters, such as, e.g.,
anomalous magnetic moments of constituent quarks or similar, have been introduced in the calculation of the e.m. $N$ structures. All results presented before
in Refs.~\cite{Wagenbrunn:2000es,Boffi:2001zb} and discussed here are pure predictions
by the GBE RCQM.

\begin{figure}
\begin{center}
\includegraphics[width=0.49\textwidth]{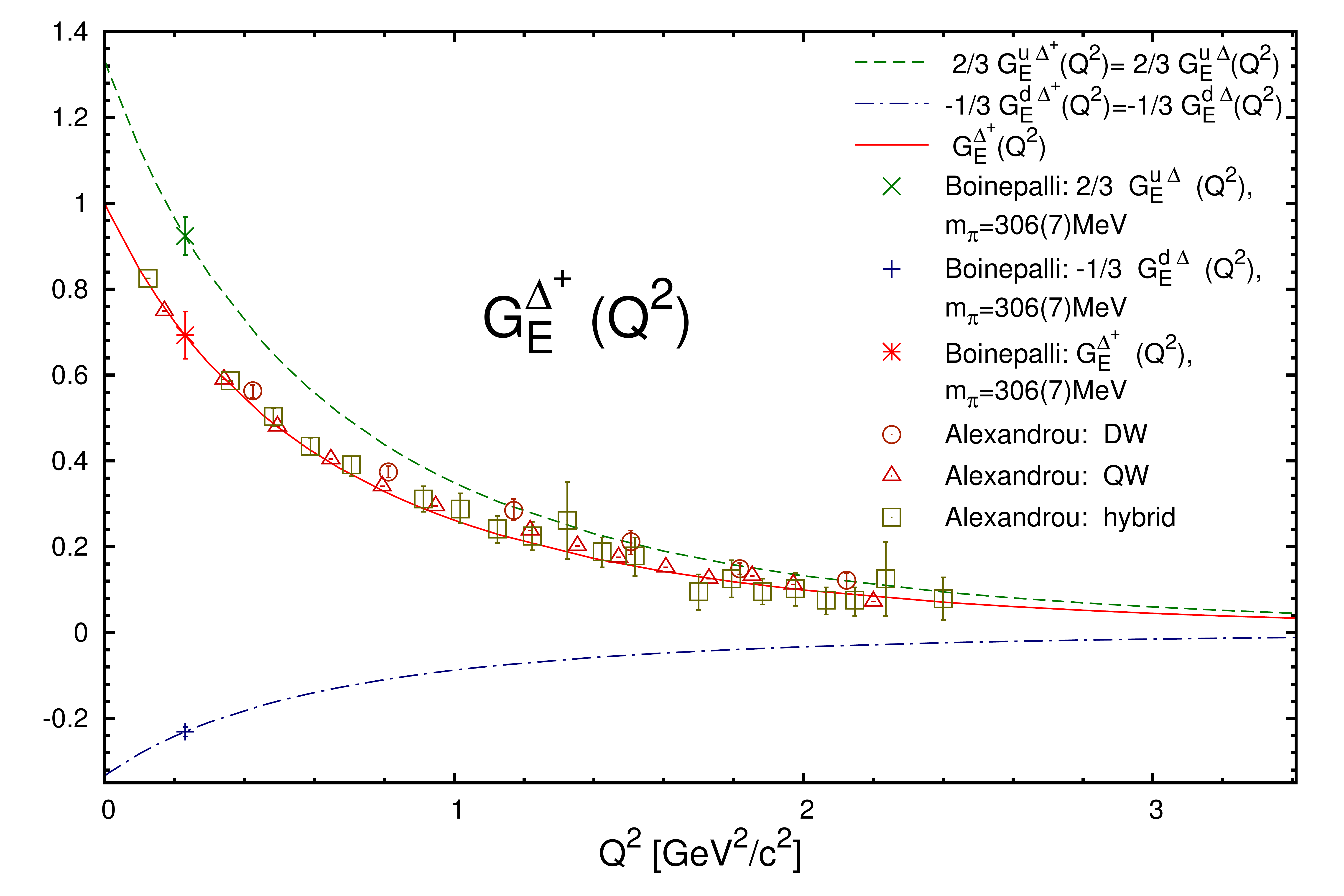}
\includegraphics[width=0.49\textwidth]{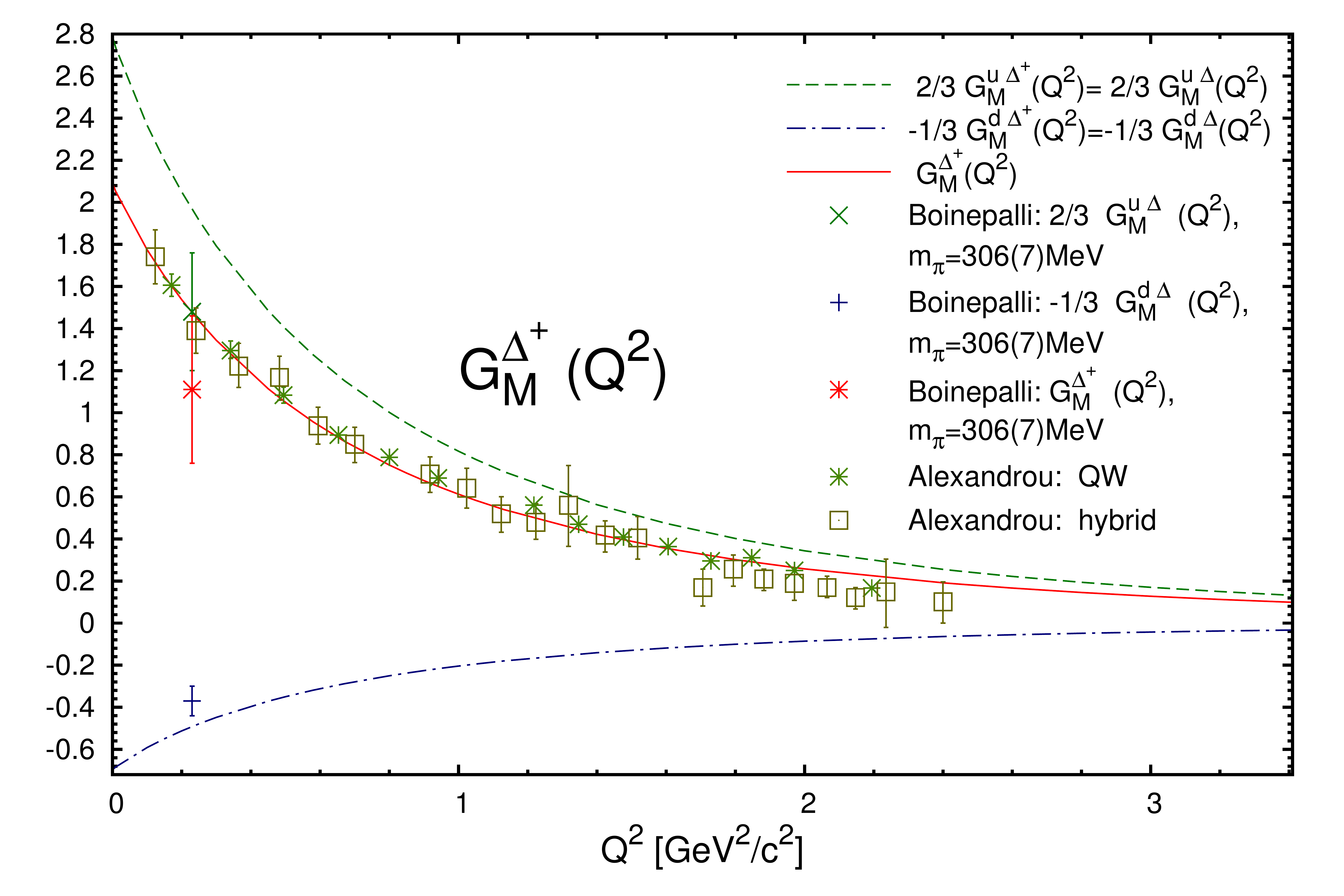}
\end{center}
\caption{Covariant predictions of the GBE RCQM for electric and magnetic form factors of the
$\Delta^+$ (solid/red lines) together with their flavor components (broken lines as specified in the inserts) in comparison to available lattice QCD
results~\cite{Alexandrou:2008bn,Boinepalli:2009sq}.} 
\label{DeltaFFs}
\end{figure}

Next we take a look at the $\Delta$'s. There is not yet any phenomenological insight into the
momentum dependences of the e.m. form factors. Experimental data exist only for the $\Delta^+$ and $\Delta^{++}$ magnetic moments. The GBE RCQM predictions for $\Delta$ and hyperon electric
radii and magnetic moments were presented in Ref.~\cite{Berger:2004yi}.
However, in the case of the $\Delta^+$ we can compare the e.m. form factors with regard to
their momentum dependences to lattice QCD results by Alexandrou et al.~\cite{Alexandrou:2008bn}
and at the point $Q^2$=0.230$\pm$0.001 GeV$^2$ also to results by Boinepalli et
al.~\cite{Boinepalli:2009sq}. We find a reasonable agreement of the covariant predictions
by the GBE RCQM with the lattice QCD results for the global form factors and at the single
point also for the flavor components in $G_E^{\Delta^+}$; there might be a discrepancy from
Ref.~\cite{Boinepalli:2009sq} for the flavor components in $G_M^{\Delta^+}$. However, for
these lattice QCD results the theoretical errors appear to be relatively big (cf. the right panel in Fig.~\ref{DeltaFFs}).

\begin{figure}
\begin{center}
\includegraphics[width=0.49\textwidth]{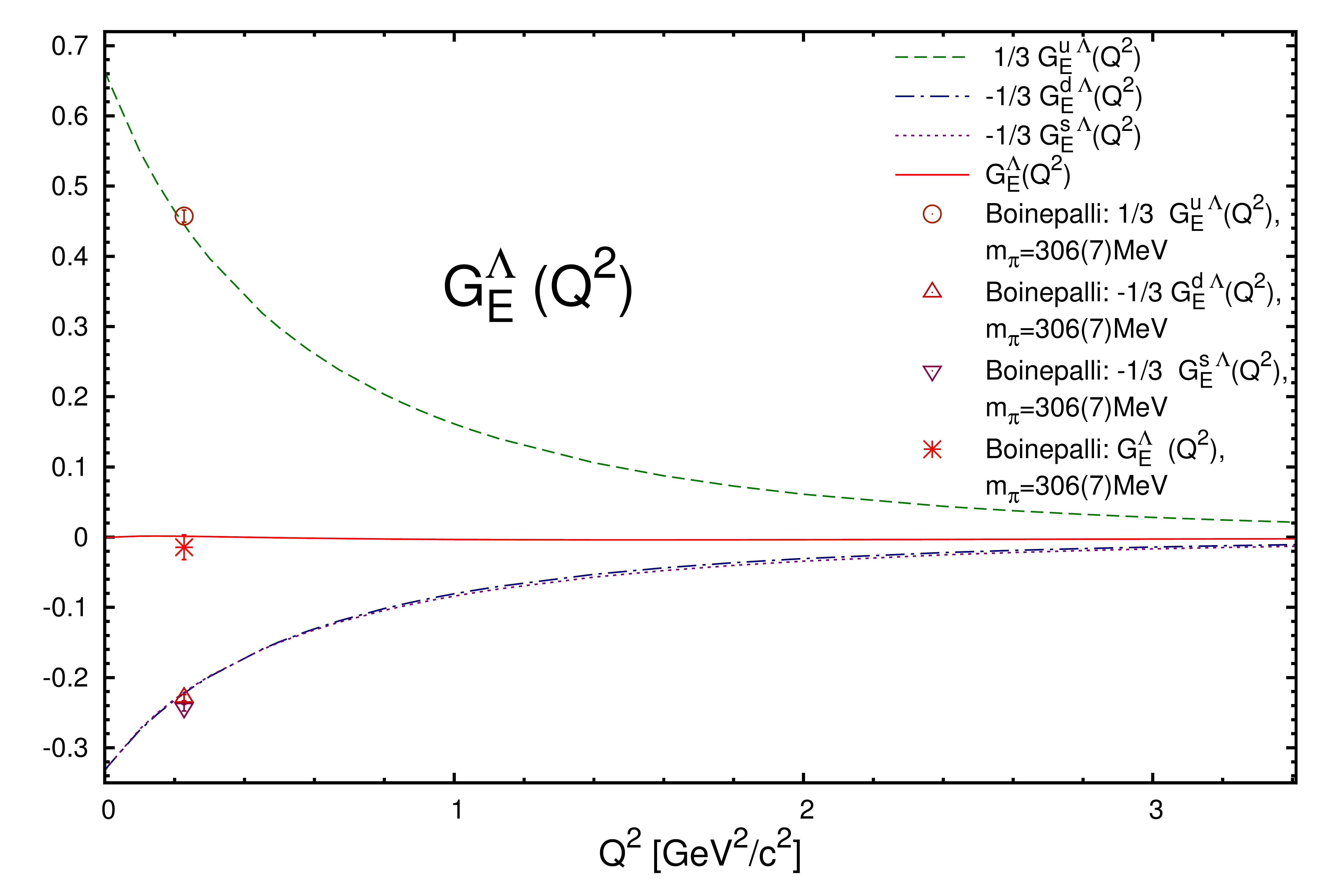}
\includegraphics[width=0.49\textwidth]{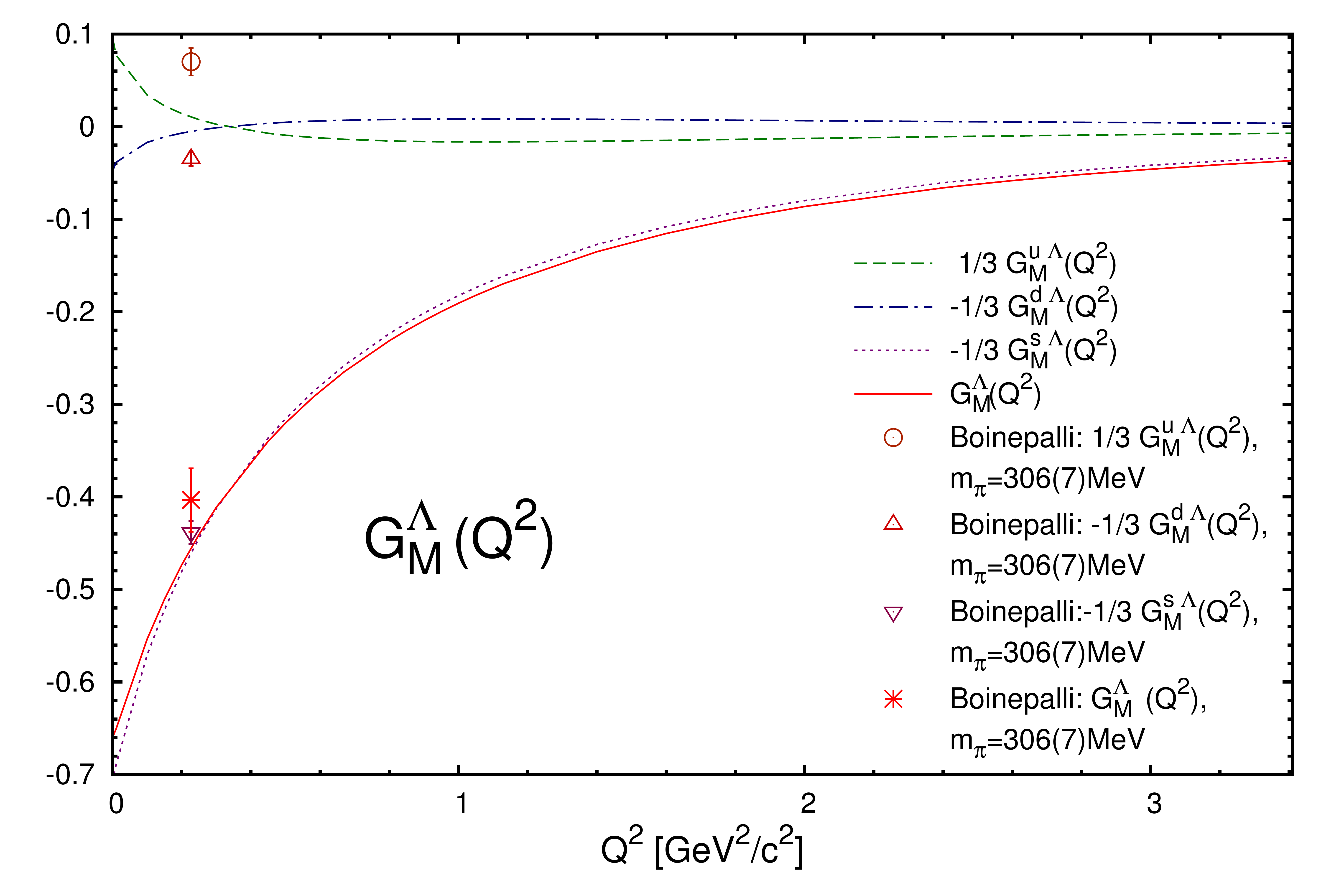}
\end{center}
\caption{Covariant predictions of the GBE RCQM for electric and magnetic form factors of the
$\Lambda^0$ (solid/red lines) together with their flavor components (broken lines as specified
in the inserts) in comparison to available lattice QCD
results~\cite{Boinepalli:2006xd}.} 
\label{LambdaFFs}
\end{figure}

For an example of hyperon e.m. form factors we here address the $\Lambda^0$ (octet) ground
state, where also an $s$ quark is involved. Fig.~\ref{LambdaFFs} shows the results for the total form factors and their flavor components. Like in the case of the $n$, the electric
$\Lambda^0$ form factor is almost zero but not quite. The main reason for this behaviour
is the small but relevant mixed-symmetry component in the spatial part of the octet wave function. Regarding the flavor components, $G^u_E$ is biggest, $G^d_E\sim G^s_E$,
and both of the latter together almost cancel with the former producing the small values of
$G^\Lambda_E$. The situation is completely different with regard to the
magnetic form factor $G^\Lambda_M$. Here, both $G^u_M$ and $G^d_M$ are extremely small and
in addition they are of opposite signs. As a consequence they have a negligible contribution
to the total magnetic form factor, which is practically only furnished by the $s$ quark
yielding $G^\Lambda_M\sim G^s_M$, a very remarkable result.
Again, as in the cases of $n$ and $\Delta^+$ the
lattice QCD results by Boinepalli et al.~\cite{Boinepalli:2006xd} for the magnetic form
factor deviate to some extent from the GBE RCQM predictions.

All $\Delta$ and hyperon e.m. form factors will be reported and discussed in a forthcoming
paper. In addition further details on the flavor decomposition of the $N$ e.m. form factors
will be given therein.

Here, we summarize only by stating that the $N$, $\Delta$, and hyperon e.m. structures are
remarkably well predicted by the GBE RCQM up to momentum transfers of $Q^2\sim$ 4 GeV$^2$.
We emphasize again that this RCQM relies on pure $\{QQQ\}$ configurations only, but
implements rigorously Lorentz symmetry
together with all other symmetry requirements of the Poincar\'e group as well as time
reversal invariance and current conservation. Obviously, also the essential properties of
low-energy QCD are grabbed well through the employed $Q$-$Q$ interaction, which is based on
a realistic (linear) confinement and a hyperfine potential that is deduced from
spontaneous breaking of chiral symmetry~\cite{Glozman:1997ag,Glozman:1997fs}.

\end{document}